\documentclass[twocolumn,showpacs,amsmath,amssymb,preprint,tightenlines,10pt]{revtex4}
\pdfoutput=1
\usepackage{graphicx,anysize}
\usepackage{bm,bbm,epstopdf}
\usepackage{multirow}
\usepackage{amsmath}
\usepackage{bbm}
\usepackage{amssymb}
\usepackage{threeparttable}
\usepackage{float}
\usepackage{varioref}
\usepackage{epic}
\usepackage{eepic}
\usepackage{makeidx}
\usepackage{epsfig}
\usepackage{subfig}
\usepackage{supertabular}
\newcommand{\diff}{\,\mathrm{d}}
\begin{document}
\preprint{To appear in Phys. Rev. E}
\bibliographystyle{prsty}

\title{ Monte Carlo simulations of fluid vesicles with in plane orientational ordering}

\author{N. Ramakrishnan }
 \email{ram@physics.iitm.ac.in}
\author{P. B. Sunil Kumar}
 \email{sunil@physics.iitm.ac.in}
\affiliation{Department of Physics, Indian Institute of Technology Madras, Chennai  600036, India}
\author{John H. Ipsen}
\email{ipsen@memphys.sdu.dk}
\affiliation{MEMPHYS- Center for Biomembrane Physics, Department of Physics and Chemistry, \\
University of Southern Denmark, Campusvej 55, DK-5230 Odense M, Denmark } 

\date{\today}
\begin{abstract}
 We present a  method for simulating fluid vesicles with in-plane orientational ordering.  The method involves computation 
 of local curvature tensor and  parallel transport of the orientational field on a randomly triangulated surface. It is shown that  
 the  model reproduces the known  equilibrium conformation of fluid membranes and  work well for a large range of bending rigidities. 
 Introduction of  nematic  ordering leads to  stiffening of the membrane.  Nematic ordering can also result in anisotropic rigidity on the surface leading to formation of  membrane  tubes.
\end{abstract}

\pacs{{PACS-87.16.D-,} {Membranes, bilayers and vesicles.} 
                           {PACS-05.40.-a} {Fluctuation phenomena, random processes, noise and Brownian motion.} 
  	                   {PACS-05.70.Np} {Interfaces and surface thermodynamics}    }

\keywords{vesicles -- membranes -- bending elasticity -- statistical mechanics --  --  membrane shape -- conformational
      fluctuations -- Monte-Carlo integration -- liquid crystals  --  -- XY model -- Lebwohl-Lasher model}

\maketitle


\section{Introduction}
The phenomenological models of fluid membrane conformations have a remarkable simplicity due to the symmetry
constraints they must obey \cite{Helfrich_1973}. However, elementary  questions on the large scale properties of
 fluid membranes  remain unresolved due to the  technical complexity in analyzing the statistical mechanics of these
 membrane models. This is in particular the case for the entropy dominated properties of membranes where assumptions
 of small configurational fluctuations or perturbative considerations fail. But even for a description of the  membrane shapes
  at the mean-field level there are many  challenges.  An alternative to the analytical approach is  computer simulations 
  of  self-avoiding fluid surfaces, which is viable both for studies of non-perturbative phenomena and shape transformations. 
  The numerical models of fluid membranes have been analyzed extensively,  in particular plaquette models, where the surface 
  is constituted by the plaquettes of a three-dimensional (3D) lattice ~\cite{Talmon_1978,  de_Gennes_1982, Duurhus_1983}, or $O(n)$ lattice gauge
   models for $n \rightarrow 0$ in 3D~\cite{Stella_1987}. A drawback  with the regular lattice based models of fluid membranes is the discrete nature of the surface configurations, which  make a detailed description of surface properties impossible and introduce phenomena which are not relevant for fluid  membranes, e.g., the roughening transition. \smallskip\\
The third class of numerical models for membranes is the triangulated random  surfaces, which were introduced in statistical mechanics in context of Euclidean string theory\cite{David_1985,Kazakov_1985,Ambjorn_1985,Polyakov_1981}. Combined  with simulation techniques for self-avoiding polymers, the triangulated random surfaces  served as  models for lipid membrane conformations~\cite{Ho_1990}.  The fluid nature of the membrane is represented by a planar, triangular lattice structure, which is allowed to change  connectivity throughout the simulation. A major advantage  of these dynamically  triangulated surface models is that discrete surface operators  can be established which posses a simple continuum limit. The results from computer simulations of randomly triangulated surfaces  can thus be interpreted in terms of continuum theory of membranes,  the related literature has been reviewed in  ~\cite{Morse_1997,Piran_2003}.  \smallskip \\
 So far  triangulated surface models only allowed  for computer simulations of membranes equipped with pseudo scalar  or scalar order parameters, e.g., mean curvature and density, while many interesting physical questions arises when vector  or tensor order parameter fields  are present in the plane of the membrane. For instance,  tilting  of the lipid molecules with respect to the surface normal,  occurring in  several of the  ordered phases of lipid bilayers, give rise to in-plane orientational ordering \cite{Nagle_2000}.  Furthermore, two good experimental evidences for the hexatic nature of the gel  phase of lipid bilayer membranes have been reported recently\cite{Bernchou_2009,Watkins_2009}. Several classes of membrane inclusions have the character
 of in-plane nematogens, e.g., antimicrobial peptides \cite{Bouvrais_2008}  and Bar domain proteins, also see \cite{Zimm_2006} and references within. In-plane orientational order in membranes  has received major attention in the theoretical literature. In particular the properties of hexatic membranes 
\cite{Nelson_1987,David_1987} and the Kosterlitz-Thouless  transition phenomena on membranes \cite{Park_1996}, the effect of lipid tilt and chirality 
\cite{ Helfrich_1988, Powers_1992, Selinger_1996,Tu_2007, Jiang_2007,Koibuchi_2008}, and the effect of surfactant polar head order \cite{Fournier_1998}.

Here we present an approach to triangulated surface models of fluid membranes by combining the
existing  simulation technique of dynamical triangulation with an approach to compute the discretized local curvature tensor.
The properties of the random surface in the new description are consistent with those from earlier models.
 
 Furthermore,  we  study  membranes  with in-plane  nematic order and show that  it can give rise to non-trivial shapes.   The paper is organized as follows:  Sec. II introduces continuum models of membranes, the Helfrich Hamiltonian and its extension to  include in-plane nematic fields with explicit  coupling to the membrane curvature.  In Sec. III we present  the  triangulated surface model which  includes a detailed description of the  local surface topography, parallel transport along the surface and our numerical implementation of the model. The Monte Carlo procedure for computer simulation of the equilibrium properties of  the triangulated surface model with in-plane orientational fields is described in Sec. IV.  In Sec. V we characterize the nature of the triangulated surface for different values of the bending moduli, without any in-plane order,  and  compare   our results  with that obtained from  earlier  simulations of membranes. In Sec. VI  we discuss some examples, in our discretized  membrane model, where the effects of the in-plane ordering  lead to some interesting shapes.        
 \section{Continuum Models} 
  It has for long time been recognized that the large scale conformations of a simple closed fluid lipid membrane can be modeled  by the Helfrich curvature energy functional \cite{Helfrich_1973}
\begin{equation}
\label{helfrichham}
{\cal H}_c = \frac{\kappa}{2}\int_s\!\! {\diff A} \,(2 M-2 C_0)^2 + \frac{\bar{\kappa}}{2}\int_s\!\! {\diff A} \, K 
\end{equation}
It is a purely geometrical model, where the characteristics of the surface is described  by the conformation of the membrane governed by the material 
 constants,  $\kappa$ the elastic bending rigidity, $\bar{\kappa}$ the Gauss curvature modulus and $C_0$ the spontaneous mean curvature. $K$ and $M$ are the local Gauss and mean curvature of the surface respectively. There are several possible extensions of Eq.(\ref{helfrichham}), e.g., describing the effects of membrane  inclusions ,  in-plane density fluctuations or in-plane order. Here we will
  discuss simple extensions of Eq.\eqref{helfrichham}, now involving in-plane vector ${\hat n}$ or a nematic tensor 
  ordering field $\frac{1}{2}({\hat n} \otimes {\hat n})$. For a vector field,  represented by an unit vector $\hat{n}$, there is only one possible relevant extension of Eq.\eqref{helfrichham}, to the lowest order in the order parameter\cite{David_1987},
\begin{equation}
{\cal H}_{vec} = \frac{K_{A}}{2} \int_s \!\!\diff A\left( \nabla {\hat n} : \nabla {\hat n} \right) \label{Eq:vector}
\end{equation}
\noindent
 which facilitates an implicit coupling of the membrane geometry to the ordering field.
 $K_{A}$ is the stiffness constant and $\nabla$ is the covariant gradient.  
The model and its extensions  have been analyzed in great detail 
(for review see chapters by Nelson,  David and by Gompper and Kroll  in \cite{Piran_2003}). 
For a nematic field, to the same order, the corresponding term is the well known Frank's free energy for nematics
 ~\cite{Frank_1958}
 \begin{equation}
\label{Eq:nem-nem}
{\cal H}_{nem}= \int_s\diff A\left\{ \frac{K_1}{2}(Div ({\hat n}))^2 + \frac{K_3}{2}(Div ({\hat n}^{\perp}))^2 \right\}.
\end{equation}
\noindent
${\hat n^{\perp}}$ is orthogonal to ${\hat n}$ in the same plane. The in-plane $Div({\hat  n})$ and $Div ({\hat n^{\perp}})$ are the  splay and bend contributions of the nematic field, and $K_1$ and $K_3$ are the corresponding Frank constants. The  in-plane nematic field gives rise to a number of new relevant couplings between the ordering field and the curvature tensor \cite{Peliti_1989}. A natural  form of the free energy,  that describes an explicit coupling between the orientational field and the curvature tensor, is given by \cite{Frank_2008, Helfrich_1988, Powers_1992,Selinger_1996,Tu_2007,Jiang_2007}
\begin{eqnarray}
\label{Eq:nem-cur}
{\cal H}_{nc} = \int_s \!\!\diff A\, \left[ \frac{\kappa_{\parallel}}{2}(H_{n,\parallel}-c_{0}^{\parallel})^2 +
              \frac{\kappa_{\perp}}{2}(H_{n,\perp}-c_{0}^{\perp})^2\right]
\end{eqnarray}
\noindent
where, $H_{n,\parallel}$ is the directional curvature along $\hat{n}$ and  $H_{n,\perp}$ is 
the directional curvature along $\hat{n}^{\perp}$. $c_{0}^{\parallel}$ and $c_{0}^{\perp}$ are  the corresponding spontaneous curvatures.
  $\kappa_{\parallel}$ and $\kappa_{\perp}$ respectively are the bending stiffness along $\hat{n}$ and $\hat{n}^{\perp}$ .


\section{Triangulated surface model}\label{exp_obs}
\begin{figure}
\begin{center}
\includegraphics[width=4.5cm]{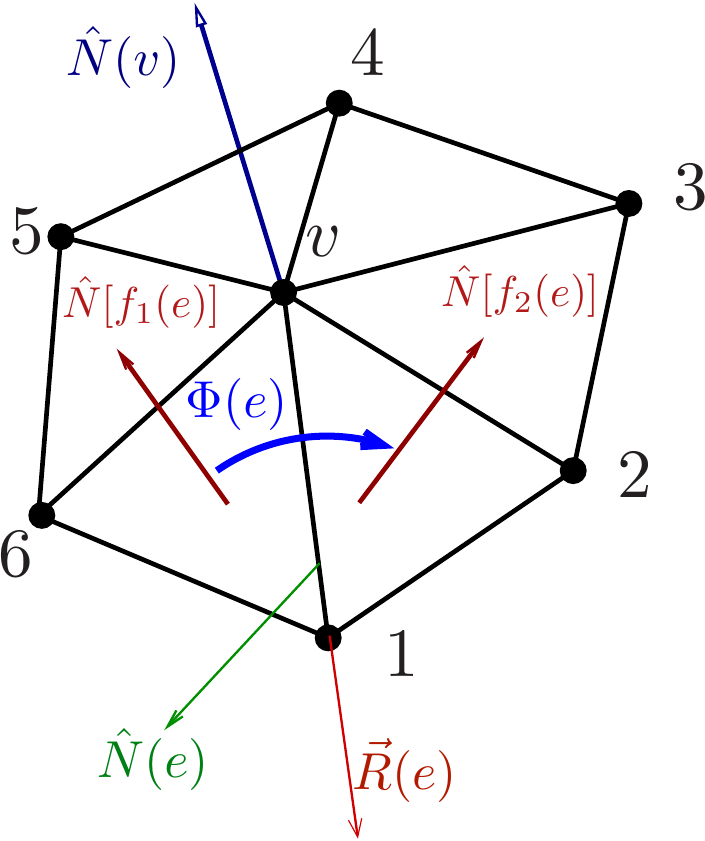}
\end{center}
\caption{(Color online) Surface patch in a one ring neighborhood around vertex $v$. The edge $e$ connects, in this description, $v$ to $1$.
 The edge vector is $\vec{R}(e)=\vec{X}(1)-\vec{X}(v)$ and  $\hat{N}(e)$ is its normal. Edge $e$ is shared by two faces $f_1(e)$ and $f_2(e)$ with $\hat{N}(f_1(e))$ and $\hat{N}(f_2(e))$, respectively, being their normals. The normal to vertex $v$ is represented by $\hat{N}(v)$.}
\label{surface}
\end{figure}

In this section we will consider discretized surfaces with the topology of a sphere, while the
considerations can readily be extended to closed triangulated surfaces 
of arbitrary topology\cite{Ipsen_1993,kroll_1998}. Contrary to the standard differential geometry  of continuum models, the discretized formulation in this section is given in Cartesian coordinates. The surface is discretized by a triangulation
 ${\cal T}^N$ consisting of $N$ vertices connected by $N_L=3 (N-2)$ links, or tethers, forming 
 closed planar graphs. The graph form a system of $N_T=2(N-2)$
 triangles corresponding to a surface with total Euler index $\chi=N-N_T-N_L=2$. 
Each vertex $v$ takes a position $\vec{X}(v)$ in $\mathbb{R}^3$. 
The triangulation and the vertex position together form  a discretized surface, 
a patch of which is given in Fig.~\ref{surface}. 
The self-avoidance of the surface is ensured by assigning a hard core spherical bead of unit diameter to each vertex and a maximal tether distance of
$\sqrt{3}$. This is in general not sufficient to impose strict self avoidance \cite{Ipsen_1995,Ho1_1990}, but a mild  constraint on the dihedral angle between two faces sharing a tether restores self avoidance. \\

The in-plane orientational field can be included by defining a unit vector $\hat{n}(v)$ in the tangent plane at each vertex $v$. In the following we will give meaning to this statement by analysis of the local surface topography and in turn calculate the curvature tensor, principal directions  and curvature invariants\cite{Polthier_2004,Polthier_2005}. The approach is based on the construction of the discretized "shape operator"  given by the differential form $-d\hat{N}$  in the plane of the surface, which  contains all information about the local surface topography. \\
 Consider a local neighborhood around a vertex  $v$, as shown in Fig.~\ref{surface}.  $\vec{R}(e)$ is the edge vector that links  $v$ to a neighboring vertex. The set of edges linked to $v$ is $\{e\}_v$, while the oriented triangles or faces with $v$ as one of their vertex is $\{f\}_v$. The calculation of the surface quantifiers at  $v$ is restricted to the one ring neighborhood around it,  which is well defined by  $\{e\}_v$ and $\{f\}_v$. Similarly the set of faces sharing an edge is given by $\{f\}_e=[f_{1}(e), f_{2}(e)]$.  The normal to an edge $e$ then is defined as,
 \begin{equation}
 \hat{N}(e)=\frac{\hat{N}[f_1(e)]+\hat{N}[f_2(e)]}{\left| \hat{N}[f_1(e)]+\hat{N}[f_2(e)] \right|},
 \end{equation}
 \noindent
where $\hat{N}[f_1(e)]$ and $\hat{N}[f_2(e)]$ are the unit normal vectors to faces $f_1(e)$ and $f_2(e)$ respectively. \\

\begin{figure}[!h]
\centering
\includegraphics[width=8.cm,height=2.5cm,clip]{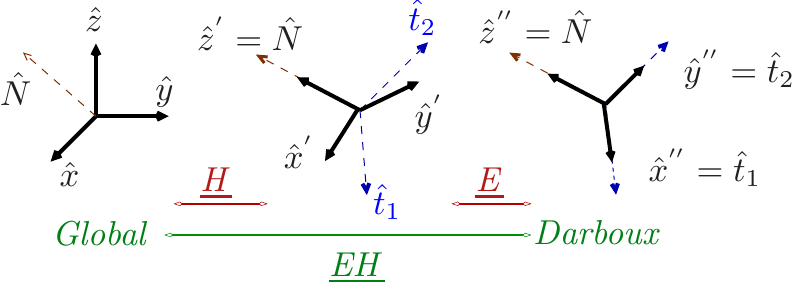}
\caption{(Color online) Transformation from a global to local coordinate frame. }
 \label{fig:hholder}
\end{figure}

We will now construct the shape operator at every vertex $v$. Toward this, we define
\begin{equation}
H(e)=2 \left | \vec{R}(e) \right| \cos \left(\frac {\Phi(e)}{2} \right).
\end{equation}
which quantifies the curvature contribution along the direction mutually perpendicular to $\vec{R}(e)$ and $\hat{N}(e)$ ~\cite{Polthier_2004,Polthier_2005, Polthier_2002}. $\Phi(e)$ is the signed dihedral angle between the faces, $f_{1}(e), f_{2}(e)$, sharing the edge $e$ calculated as
\begin{equation}
\begin{split}
\Phi(e)  =  {\rm sign}\left[ \left\{{\hat N}[f_1(e)]\times {\hat N}[f_2(e)] \right\}\cdot  \vec{R}(e)\right] \\ 
 \arccos \left[{\hat N}[f_1(e)] \cdot {\hat N}[f_2(e)] \right] + \pi. 
 \end{split}
\end{equation}
The discretized ``shape operator'', which quantifies both the curvature and the orientation of $e$ is thus  the tensor
\begin{equation}
\underline {\mathbf{S_e}}(e) = H(e) \left [{\hat R(e)} \times {\hat N}(e)\right ] \left [{\hat R(e)}\times {\hat N}(e)\right ],
\label{eq:shape_e}
\end{equation}
where $\hat{R}(e)=\vec{R}(e)/| \vec{R}(e)|$ is the unit vector along edge $e$. Having defined the shape operators, \{$\underline{\bf S_e}(e)$\}, along the edges 
of the vertex $v$, we now proceed to compute the shape operator at $v$. The normal to the surface at $v$ can be calculated as,
\begin{equation}
 \hat{N}(v)= \frac{\sum_{\{f\}_v} \Omega[A(f)]\,\hat{N}(f) }{\left|\sum_{\{f\}_v}  \Omega[A(f)]\,\hat{N}(f)  \right| },
\end{equation}
\noindent
with $A(f)$ denoting the surface area of the face $f$ and the normalized weight factor $\Omega[A(f)]$ is proportional to the area of the face. The projection operator, $\underline{{\bf P}}(v) = \mathbbm{1} - \hat{N}(v) \hat{N}(v)$, 
projects  \{$\underline{\bf S_e}(e)$\} on to the tangent plane at $v$~\cite{Polthier_2004,Polthier_2005}. The shape operator at the vertex $v$ is then a weighted sum of these projections given by 
\begin{equation}
{\underline {\bf S_v}(v)} = \frac{1}{A(v)}\,\,\sum_{\{e\}_v} W(e)\,\underline{ {\bf P}}(v)^{\dagger}\,\underline{{\bf S_e}}(e)\,\underline{{\bf P}}(v).
\label{eq:curv_operator}
\end{equation}
$A(v)=\sum_{\{f\}_v} A(f)/3$ is the average surface area around $v$, while the weight factor for an edge is calculated as $W(e)= \hat{N}(v) \cdot \hat{N}(e)$. The shape operator Eq.\eqref{eq:curv_operator} at the vertex $v$ is  expressed in coordinates  of the global reference system $[\hat{x},\hat{y},\hat{z}]$.  Notice that, by construction,  the vertex normal $\hat{N}(v)$ is an eigenvector of, $\underline{\mathbf{S_v}}(v)$, corresponding to eigenvalue zero.  The two other principal directions $\hat{t}_{1}(v)$, $\hat{t}_{2}(v)$, whose corresponding eigenvalues are the principal curvatures, define the tangent plane at the vertex $v$.  A local coordinate frame called the Darboux frame $[\hat{t}_{1}(v),\, \hat{t}_{2}(v),\, \hat{N}(v) ]$, see Fig.\ref{fig:hholder}, can then be defined at $v$.
\noindent
The transformation from the global to Darboux frame, see Fig. \ref{fig:hholder}, is obtained by first applying a Householder transformation($\underline{\mathbf{H}}$), see Appendix A, to rotate the global $\hat{z}$ direction into $\hat{N}(v)$, while $\hat{x}$ and $\hat{y}$ are rotated into vectors ${\hat x}', {\hat y}'$ in the tangent plane at the vertex $v$.  The shape operator, at $v$,  in this frame $\underline {{\bf C}}(v)=\underline {{\bf H}^{\dagger}}(v)\,\underline {{\bf S_v}}(v)\,\underline {{\bf H}}(v)$ is a 2x2 minor, with the two principal curvatures $c_{1}(v)$ and $c_{2}(v)$ as its eigenvalues. The corresponding eigenvector matrix $\underline{\bf E}(v)$  transform $[\hat{x}^{'},\hat{y}^{'},\hat{N}(v)]$  into the Darboux frame at $v$. Any vector in the global frame, can now be transformed to this local frame by the transformation matrix $\underline{{\bf E}}(v)\,\underline{{\bf H}}(v)$\smallskip .
  
We are now in the position to write up the discretized form of Helfrich's free energy, at a vertex $v$, based on the local curvature invariant 
 $M(v)=[c_{1}(v)+c_{2}(v)]/2$ and $K(v)=2 c_{1}(v) c_{2}(v)$:
\begin{equation}
 {\cal H}_c = \sum_{v=1}^N A(v)\left[\frac{\kappa}{2}\,(c_{1}(v)+c_{2}(v))^2 + \bar{\kappa}\, c_{1}(v) c_{2}(v) \right] .
\label{eq:discrhelfric}
\end{equation}
\noindent
The calculation of the discrete curvature tensor has been performed by other  methods \cite{Taubin_1995,Shimsoni_2003}, however we find the method used in this paper to be the most accurate in describing surfaces with prescribed geometry. 
The local  Darboux frame  is very useful for the characterization of an in-plane vector field $\hat{n}(v)$.
 For convenience, we choose $c_1(v)$ to be the maximum principal curvature and $\hat{t}_1(v)$ the corresponding principal  direction.  The local orientational angle $\varphi(v)$ of $\hat{n}(v)$  will always refer to  this Darboux frame.  \\
 In order to compare the orientation of two distant in-plane vectors at the surface, it is necessary to perform parallel transport of the vectors on the discretized surface. In practice we need only to define the parallel transport between neighboring vertices, i.e. a transformation  $\hat{n} (v^{'}) \rightarrow \underline{\bf \Gamma}(v,\,v^{'}) \hat{n}(v)$, which brings $\hat{n}(v)$ correctly into the tangent plane of the vertex $v^{'}$, so that its angle with respect to the geodesic connecting $v$ and $v^{'}$ is preserved. 
If ${\hat r}(v,v^{'})$  is the unit vector  connecting   a vertex $v$ to its neighbor $v^{'}$ and ${\vec \zeta}(v)$=$\underline{{\bf P}}(v){\hat r}(v,v^{'})$  and ${\vec \zeta}(v^{'})$= $\underline{{\bf P}}(v^{'}) {\hat  r}(v^{'},v)$ are its projection on to the tangent planes at $v$ and $v^{'}$; then  our best estimate for  the directions of the geodesic  connecting them,  are the unit vectors ${\hat  \zeta}(v),\,{\hat \zeta}(v^{'})$. The decomposition of $\hat{n}(v)$ along the  orientation of the geodesic and its perpendicular in the tangent plane of $v$ is thus: 
\begin{equation}
\begin{split}
 \hat{n}(v) =  \left[\hat{n}(v) \cdot \hat{\zeta}(v)\right] \hat{\zeta}(v)+  \\
                     \left[\hat{n}(v)\cdot (\hat{N}(v) \times \hat{\zeta}(v))\right] \left(\hat{N}(v) \times \hat{\zeta}(v)\right)
\end{split}
\end{equation} 
\noindent
Parallelism now demand that these  coordinates,  with respect to the geodesic orientation, are the same  in the tangent 
plane of $v^{'}$, therefore:
\begin{equation}
\begin{split}
\underline{{\bf \Gamma}}(v,v^{'})\hat{n}(v)=  \left[\hat{n}(v) \cdot \hat{\zeta}(v)\right] \hat{\zeta}(v^{'})+ \\
 \left\{\hat{n}(v) \cdot (\hat{N}(v) \times \hat{\zeta}(v))\right\} \left[ \hat{N}(v^{'}) \times \hat{\zeta}(v^{'})\right]
\end{split}
\end{equation} 
\noindent
This parallel transport operation allow us to define the angle $\phi({v,v^{'})}$ between vectors in the tangent plane at neighboring vertices, and in turn their cosine and sine as:
\begin{eqnarray}
 \cos(\phi(v,v^{'}))&= & \hat{n}(v^{'})\cdot \underline{{\bf \Gamma}}(v,v^{'})\hat{n}(v) ;\\
 \sin(\phi(v,v^{'}))&=& \left[\hat{N}(v^{'})\times\hat{n}(v^{'}) \right] \cdot \underline{{\bf \Gamma}}(v,v^{'})\hat{n}(v) \nonumber
\end{eqnarray}
\noindent
We can now define the lattice models, corresponding to Eqs.\eqref{Eq:vector} and \eqref{Eq:nem-nem}, for the in-plane orientational field, e.g., the XY-model on a random surface:
\begin{equation}
 {\cal H}_{\rm XY}= -\frac{\epsilon_{\rm XY}}{2}\sum_{\langle vv^{'} \rangle} \cos[\phi(v,v^{'})]
\end{equation}
\noindent
or the Lebwohl-Lasher model on a random surface:
\begin{equation}
 {\cal H}_{\rm LL}= -\frac{\epsilon_{\rm LL}}{2}\sum_{\langle vv^{'} \rangle}\left\{\frac{3}{2}\cos^2(\phi(v,v^{'}))-\frac{1}{2}\right\}
\label{Eq:lebwohl}
\end{equation}
\noindent
Furthermore, we are now in a position to calculate, at a given vertex $v$, the directional curvatures along  and perpendicular
to the orientation of the in plane vector field ${\hat n}(v)$ by use of Gauss formula:
\begin{eqnarray}
M(v)_{\parallel} &=& c_1(v) \cos^2[\varphi(v)]+c_2(v) \sin^2[\varphi(v)] \nonumber\\
M(v)_{\perp} &=& c_1(v) \sin^2[\varphi(v)]+c_2(v) \cos^2[\varphi(v)] \nonumber \\
\label{eq:sectcurvature}
\end{eqnarray}
 \section{Monte Carlo procedure}
\begin{figure}
\begin{center}
\centering
\includegraphics[width=8.cm,clip]{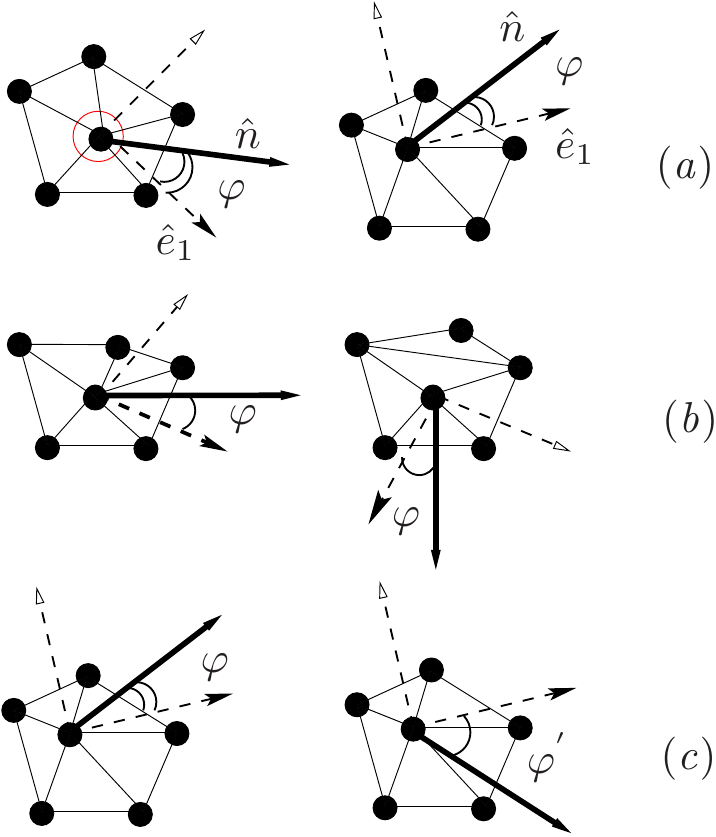}
\caption{ (Color online) Monte Carlo moves, a) vertex shift, b) link flip and c) angle rotation. Surface vector field, $\hat{n}$, is represented with  solid arrow while principal directions $\hat{e}_1$ and $\hat{e}_2$  are marked with dotted arrows. $\varphi$ is the angle $\hat{n}$ subtends with $\hat{e}_1$.}

 \label{move_classes}
 \end{center}
\end{figure}

The equilibrium properties of the discretized surface can
now be evaluated from the
analysis of the total partition function, i.e., the sum of Boltzmann factors for all surface configurations and triangulations.
For simplicity, we consider the situation with just one in-plane orientational $\hat{n}(v)$ field defined at each vertex
\begin{eqnarray}
Z(N,\kappa,\bar{\kappa},\epsilon,..)=\frac{1}{N!} \sum_{{\cal T}^{N}} \prod_{v=1}^{N} \int d\vec{X}(v) \int  d\varphi(v)\nonumber \\
\exp\left(-\beta \left({\cal H}^c(\{\vec{X}\},{\cal T}^{N},\{\varphi\})+ U_{SAS}\right )\right) \,\,\,\,\,
\label{eq:Z}
\end{eqnarray}
\noindent
where  $U_{SAS}$  is the potential that ensures the self-avoidance of the surface and $\varphi(v)$  is integrated over the unit circle or half unit circle for the XY field and the nematic field respectively.  $\{\vec{X}\}$ and $\{\varphi\}$ are, respectively, the complete set of vertex positions and orientational angles. Further, we set $\beta=\frac{1}{k_B T}=1$. In practice, a surface configuration is represented by
a tuple $\eta=(\{\vec{X}\},{\cal T}^{N},\{\varphi\})$, which must be updated during the Monte Carlo simulation procedure. The Monte Carlo updating scheme can
 now be decomposed into three move classes, so each of the three sets of degrees of freedom are updated independently to keep it simple and ensure fulfillment
 of detailed balance: \medskip \\
\noindent
{\bf Vertex shifts:} represent the updates of the vertex positions, keeping ${\cal T}^N, \{ \varphi\}$ fixed,  thus allowing for shape changes of the membrane. The attempt probability to change to a new configuration  $\eta'=(\{\vec{X'}\},{\cal T}^{N},\{\varphi\})$, with a chosen vertex moved to a new position within a cube of 
side $2\sigma$ centered around its old position, is   
$\omega(\eta|\eta') = \omega(\eta'|\eta) =([2\sigma]^3 N)^{-1}$.
$\sigma$ is appropriately chosen to get a reasonable acceptance rate of 30--50\%. In our simulations  $\sigma=0.1$
is chosen.With this surface updating operation, the curvature tensor and thus the principal axis changes.
Since the angle $\{\varphi\}$ is kept fixed, the set of orientations $\{ \hat{n}\}$, in the global frame, are changed following the local surface
configuration, Fig. \ref{move_classes}(a). \medskip \\
\noindent
{\bf Link flip:} represents  updating of the triangulation. Here a link, $e$ connecting a vertex $v$ to $v^{'}$, is picked at random and an attempt is made to flip it 
to the pair of opposite vertices common to  $v$ and $v^{'}$.  The attempt probability to change to  configuration $\eta'=(\{\vec {X}\},{\cal T}'^N,\{\varphi \})$ is then 
 $\omega(\eta|\eta') = \omega(\eta'|\eta) = 1/N_L$.  Similar to vertex shifts, the actual orientations $\{\hat{n}\}$ are now changed, following the local surface configuration, Fig. \ref{move_classes}(b). \medskip \\ 
\noindent
{\bf Angle rotation:}  the orientation of the  in-plane vector $\hat{n}(v)$, at a randomly chosen vertex $v$,  is updated. 
The vector is rotated to a new, randomly chosen,  direction in the tangent plane, keeping the vertex positions and link directions fixed. As a result of which the orientational angle is now $\varphi^{'}(v)= \varphi(v)+\Delta \varphi(v)$. The attempt probability to
 configuration $\eta'=(\{\vec{X}\},{\cal T}^{N},\{\varphi^{'}\})$ is 
$\omega(\eta|\eta^{'}) = \omega(\eta'|\eta)=(2 \sigma_{\varphi} N)^{-1}$, where $\sigma_{\varphi} \ll  \pi$ is the maximum increment  of the angle. The surface topography is not affected by this move, Fig. \ref{move_classes}(c). \\

For each of the above moves, the acceptance probability is:
\begin{equation}
acc(\eta|\eta') = {\rm min}(1,\frac{\omega(\eta^{'}|\eta)}{\omega(\eta|\eta^{'})}\exp(-\beta \left( H(\eta^{'})-H(\eta )\right))
\label{eq:15}
\end{equation}
\vspace*{3pt}
The duration of a Monte Carlo simulation is measured in MCS (Monte Carlo sweeps per Site), which represents $N$ attempted {\em vertex moves}, 
$3 (N-2)$ attempted {\em flips} and $N$ attempted  rotations of ${\hat n}$.

\section{Results and Discussion} \label{Results}
\subsection{Vesicles with no in-plane order}
\begin{figure}[!h]
\centering
\includegraphics[width=7.5cm,height=5.cm,clip]{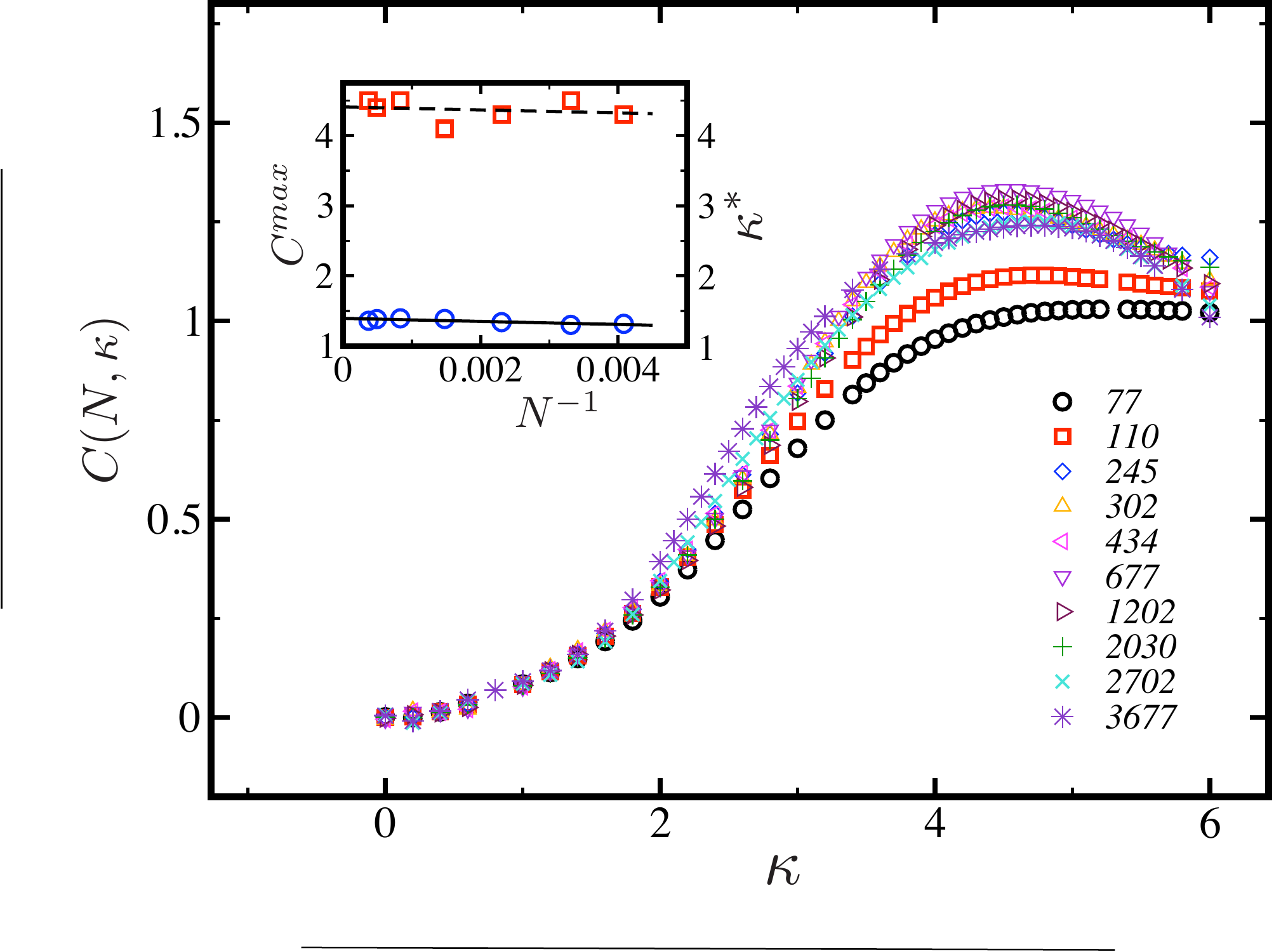}
\caption{ (Color online) Specific heat $C(N,\kappa)$ versus $\kappa$ for  varying $N$. The position of $C^{\rm max}(N,\kappa)$(\emph{circles fitted with solid line})
 and $\kappa^{*}(N)$(\emph{squares fitted with dotted line}) are shown in the inset.}
 \label{fig:C}
\end{figure}

\begin{figure}[!h]
\centering
\includegraphics[width=7.5cm,clip]{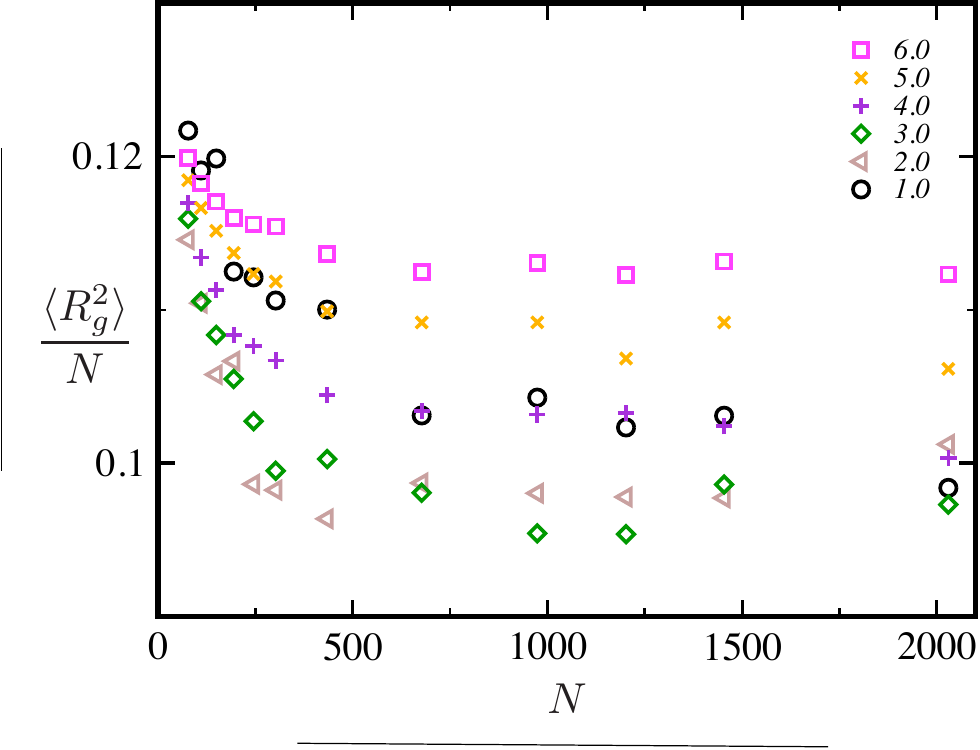}
\caption{(Color online) linear scaling of $R_g^2$ as a function of N for various $\kappa$ is shown. Entropic domination in lower $\kappa$ regime brings 
in a large spread in the values of $R_g^2/N$ for $\kappa <1.0$ and are not shown here. }
\label{fig:Rgplot}
\end{figure}

\begin{figure}[!h]
\centering
\includegraphics[width=8.0cm,clip]{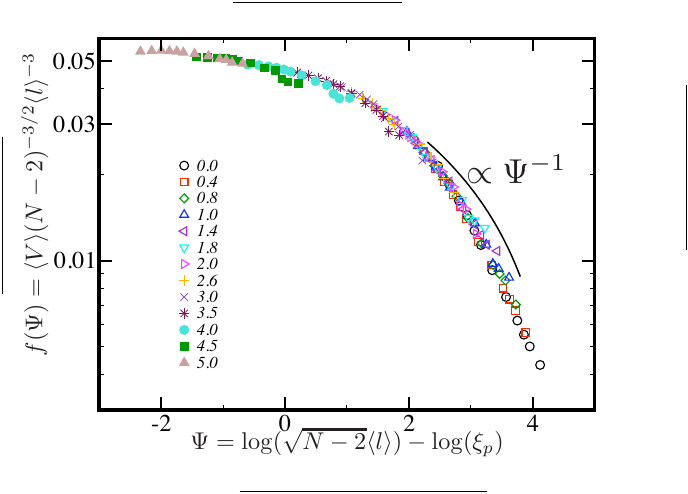}
\caption{(Color online) Universal scaling function describing the dependence of volume on the system size. The data collapse is obtained by
  determining  $\xi_p$ for each $\kappa$ separately. }
\label{fig:scaling}
\end{figure}

\begin{figure}[!h]
\centering
\includegraphics[width=7.cm]{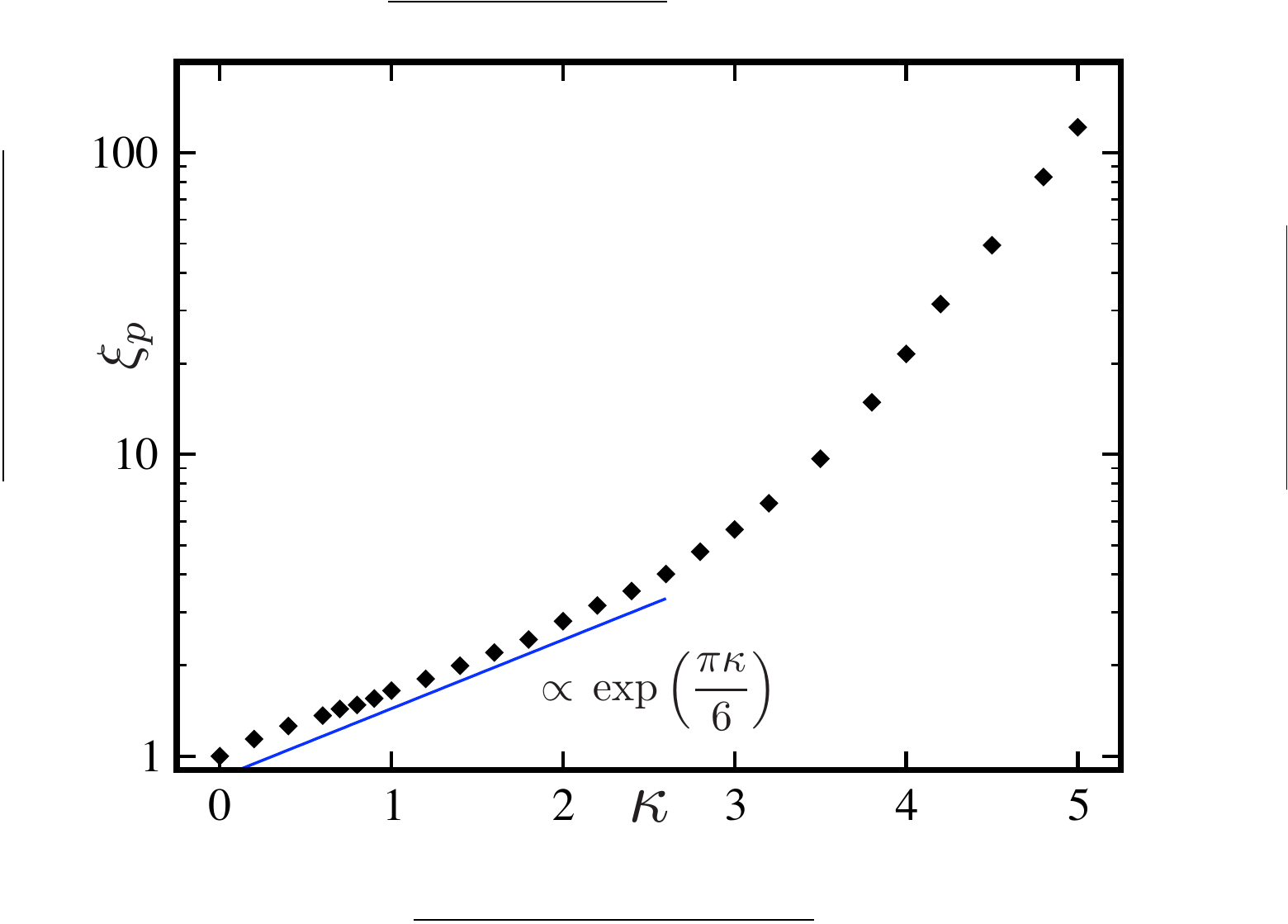}
\caption{(Color online) Persistence length, $\xi_p$, as a function of $\kappa$}
\label{fig:persleng}
\end{figure}

In the first part of this section we will discuss the properties of this new discretized random surface description of membrane 
conformations for a simple, closed, fluid membrane of spherical topology, with no in-plane order.  All simulations reported in this paper are carried out with $\overline{\kappa}=0$. System sizes in the range $N=77$ to $3677$ and  bending rigidity in the range  $\kappa = 0$ to $1000$ were investigated to  compare it with the previously known results for these systems.
\begin{figure}[!h]
\centering
\includegraphics[width=7.5cm]{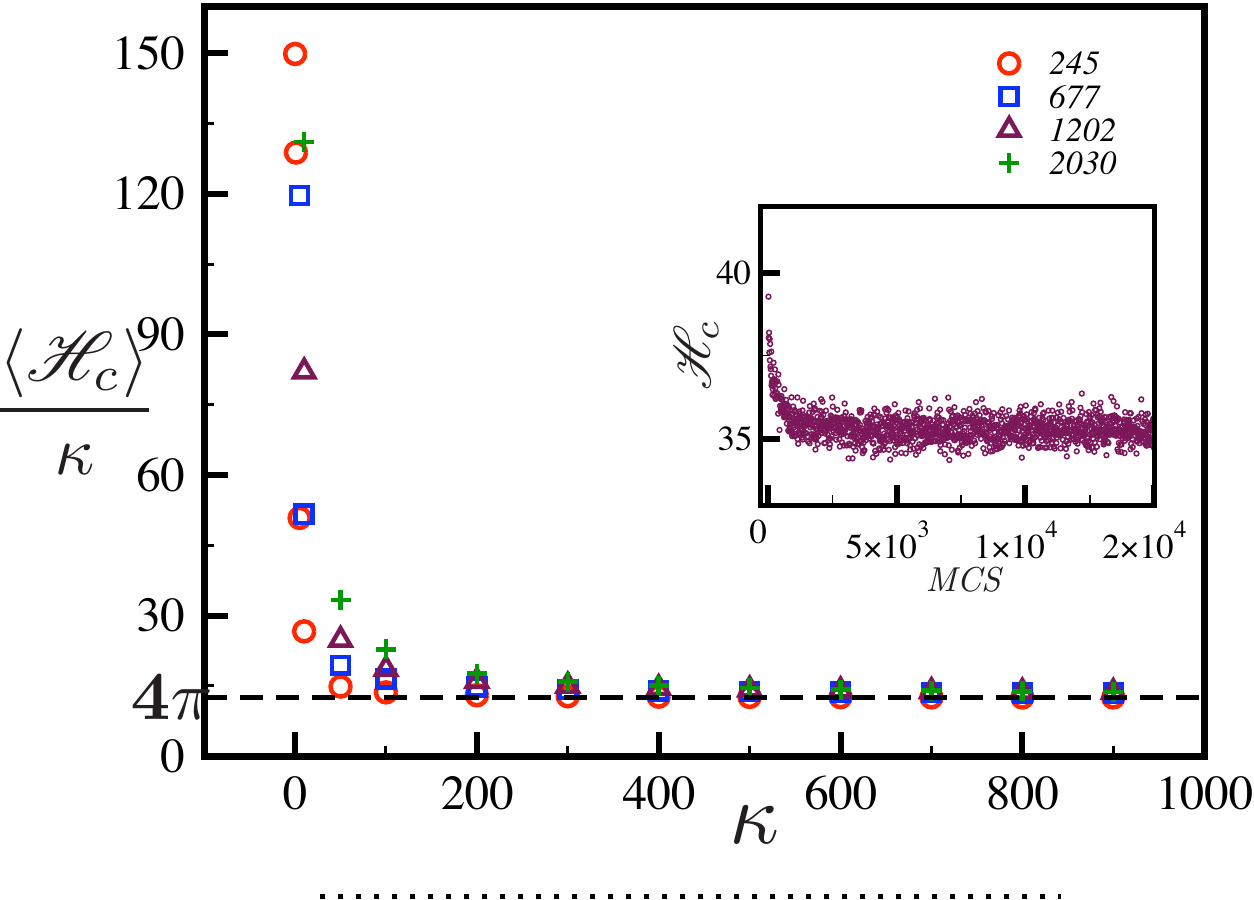}
\caption{(Color online) $\frac{\langle {\cal H}_c \rangle}{\kappa}$ versus $\kappa$ for stiff membranes
 and varying $N$.  Note that for large values of $\kappa$ the line approaches  $4 \pi$,  the value  of  $\langle {\cal H}_c \rangle$  over bending modulus, for 
a smooth sphere with bending modulus $\kappa$. Inset shows a time series of the curvature energy for system of $N=2030$ vertices for $\kappa=200$.}
\label{fig:stiff_membranes}
\end{figure}
  Applying the equipartition theorem to Gaussian or quasi-spherical configurational fluctuations  shows that the expected behavior is
$\frac{\langle {\cal H}_c \rangle}{\kappa} \longrightarrow 8 \pi + \frac{N-1}{2}\frac{1}{\kappa}$. In Fig.~\ref{fig:stiff_membranes}, it is shown  that the ensemble 
averaged curvature energy of the vesicle,  $\frac{\langle {\cal H}_c \rangle}{\kappa}$ indeed approaches $8 \pi$ for large $\kappa$.  In the opposite limit of small $\kappa$ the literature is largely focused on the  crumpling transition. 	 Such a transition should be  indicated by the presence of  a peak or a cusp in the 
specific heat, 
\begin{equation}
 C(N,\kappa) = \frac{1}{N}(\langle {\cal H}_c^2 \rangle - \langle {\cal H}_c \rangle^2).
\end{equation}
$C(N,\kappa)$, as a function of   $\kappa$ for different $N$, is shown in Fig.~\ref{fig:C}.  The shape of the curve is similar to what has been
 reported by earlier simulations \cite{Kroll_1992,Ambjorn_1993, KA_1993}.  As reported in these papers, we find that  the peak height ($C^{max}$)  stops growing 
 and the peak   position ($\kappa^*$)  saturates to a constant value beyond system size $N\approx 500$ ( see Fig.~\ref{fig:C}). In the aysmptotic limit $\kappa^*$ and $C^{max}$, in dimensionless units, saturates to  approximately 4.4  and 1.4 respectively.
The smooth and finite nature  of $C(N,\kappa)$ for large $N$ shows that this measure does not indicate the presence of a first order or a continuous transition in the thermodynamic
 limit.  However, a continuous transition cannot  be completely ruled out.  If $\kappa$ is an {\em irrelevant} thermodynamic variable under RG transformation,
 it just leaves a cusp in $C(N,\kappa)$ at the transition, a similar phenomena is well-known for the $\lambda$-transition of He$^3$-He$^4$ 
mixtures \cite{Prokrovsky_1965}.  Note that the value of $\kappa^*$ appears to be roughly five times that of the previously reported 
values\cite{Kroll_1992,KA_1993}.   This is a clear indication of that the new measure of local mean curvature differs  from that used  previously,
although the prediction of a low $\kappa$ cusp in $C(N,\kappa)$ persists. 
A simple quantifier of membrane conformations used in triangulated surface simulations is the gyration tensor 
\begin{equation}
{\underline{\bf G}} = \frac{1}{2 N^2} \sum_{v,v^{'}}^N (\vec{X}(v) -\vec{X}(v^{'}) ) (\vec{X}(v)-\vec{X}(v^{'}))^{\dagger}, \nonumber
\end{equation}
 \\with $ R^2_G =  {\rm Tr}({\bf G})$
 as the simplest invariant. For the flexible, tethered, self-avoiding random surfaces  $R_G^2\sim N^{\alpha}$. Earlier simulations report 
  $\alpha=0.8$~\cite{Ho1_1990} and  $\alpha=1.$~\cite{Kroll_1992}.  As shown in Fig. \ref{fig:Rgplot} we find that   $R_G^2 \propto N$  for all 
 values of $\kappa$, which is characteristic of the self-avoiding branched polymer and quasi spherical configurations \cite{Duurhus_1983}.  
  The similarity of the exponent makes an analysis of the cross-over, between the branched polymer configurations at  low $\kappa$  and  quasi-spherical shapes at high $\kappa$, very difficult by use of $R^2_G$. This is better accomplished by analysis of the vesicle volume, which in previous vesicle simulations have been shown 
to obey the simple scaling  ansatz $V=N^{\frac{3}{2}}f\left[\sqrt{a N}/\xi_p(\kappa)\right]$, where $f(x)$ is a scaling function and $\xi_p(\kappa)$ is a cross-over
length scale, identified with the persistence length~\cite{Kroll_1995, Ipsen_1995}.   This universal scaling  behavior also holds for our new triangulated surface
model as shown in Fig.~\ref{fig:scaling}.
 Here, for each $\kappa$, $\xi_p$ is chosen such that  we  obtain good  data collapse.  It has been found by RG-analysis that $\xi_p$  for a  fluctuating smooth continuous surface, embedded in 3D space, depends on  $\kappa$  as $\exp(4\pi\kappa /3K_bT)$ \cite{Peliti_1985}. This dependence has been verified  numerically by previous triangulated surface simulations~\cite{Ipsen_1995}.  
However, the persistence length, obtained from the scaling plots shown in Fig.~\ref{fig:scaling},  predicts  a different dependence 
on $\kappa$ ( see Fig.~\ref{fig:persleng}).  In the flexible regime,  $\kappa \leq 3 k_b T$,  an approximate exponential behavior $\exp(c \kappa /k_bT)$, $c \simeq \pi/6$ is seen,  while in the semi-flexible regime, $\kappa \geq 3 k_b T$, a stronger dependence  of $\xi_p$ on $\kappa$ is found. Our data does not allow for a determination of the asymptotic behavior of $\xi_p(\kappa)$ for large $\kappa$.  The scaling function $f(\Psi)$,  where  $\Psi=\sqrt{N}\xi_p^{-1}$,  is a constant for small $\Psi$ ( semi-flexible regime )  and is  $ \sim\,\Psi^{-1}$ for large values 
 of $\Psi $,  indicating a  branched polymer behavior in the flexible regime, see Fig.~\ref{fig:scaling}. 
This  suggest  that, in this model,  the effective bending rigidity is a decreasing function of temperature, with $c$  saturating to $4 \pi /3$ at low  temperatures.
Overall, we have shown in this section that the new algorithm reproduce the expected behavior of vesicles governed by Helfrich's free energy , given in Eq.\eqref{helfrichham}, in the rigid regime of high $\kappa$. In the flexible to semi-flexible regimes of low $\kappa$ values, our new numerical representation of the geometry and energetics of vesicles produce a behavior which is qualitatively in agreement with previous triangulated surface models of vesicles. 
 However, the cusp in the  specific heat has shifted to higher $\kappa$ value. The flexible regime at $\kappa$ values
below the cusp is more rigid compared to previous models with an approximate exponential dependence between the persistence length and $\kappa$ and $\xi_p(\kappa^*) \simeq 10$. Above $\kappa^*$ the increase in $\xi_p$ is much stronger. We attribute the differences between the present model and previous models to the use of different surface quantifiers. 
\begin{figure*}[t]
\centering
\subfloat{\label{fig:zerosp-f1}\includegraphics[width=4.cm,clip]{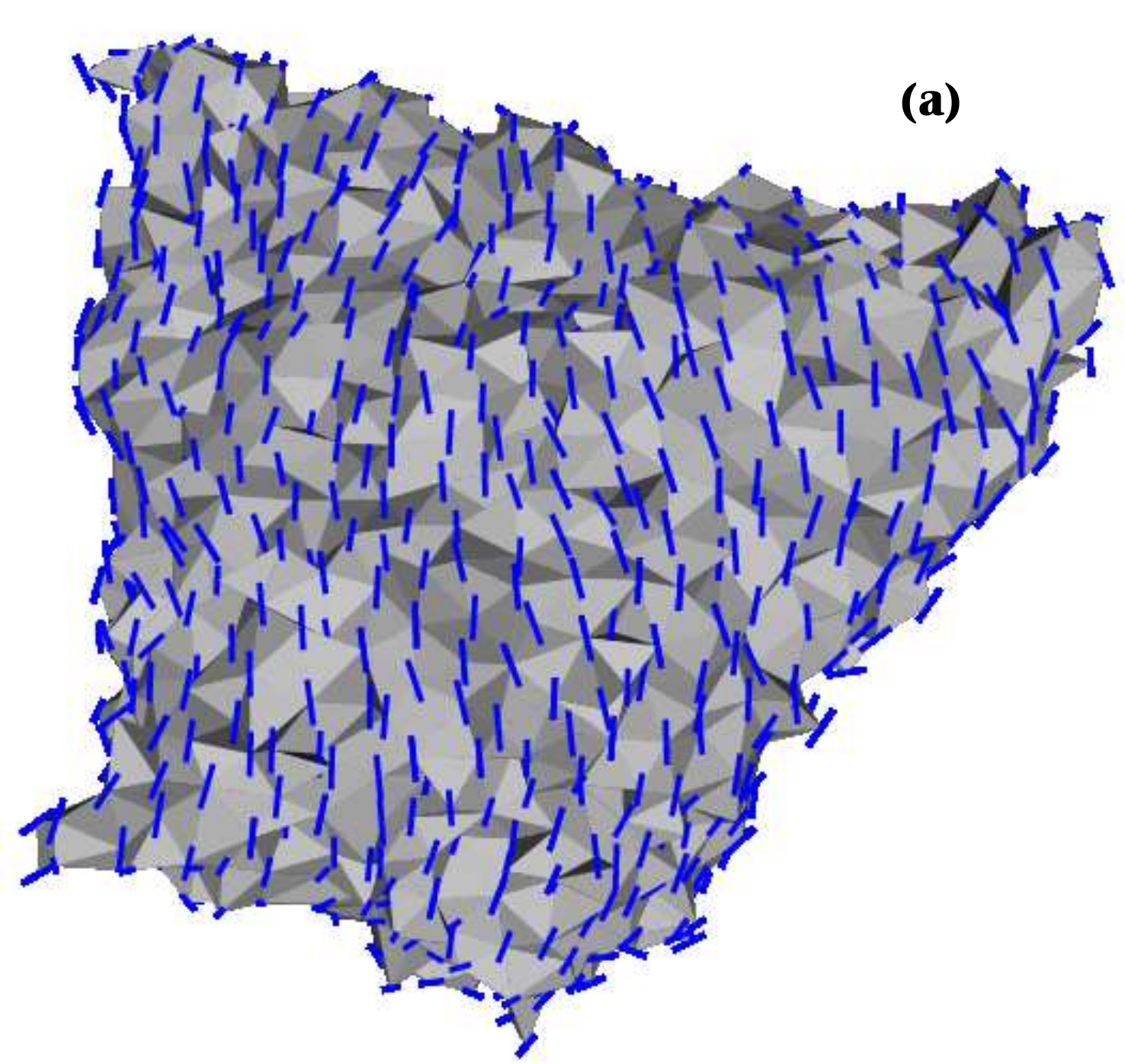}{{\large{\textbf{ (a)}}}}}
\subfloat{\label{zerosp-f2}\includegraphics[width=4.cm,clip]{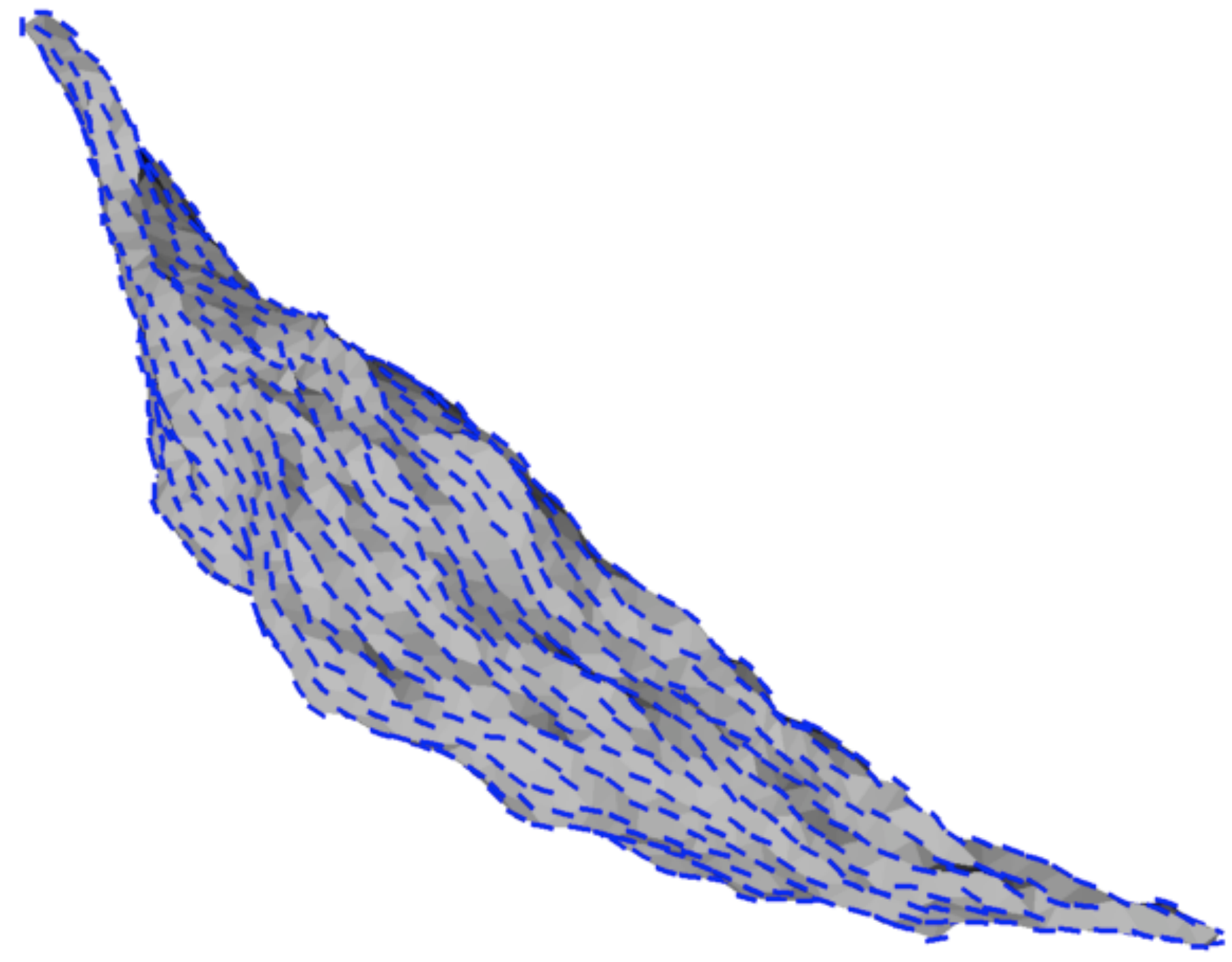}{{\large{\textbf{ (b)}}}}}
\subfloat{\label{zerosp-f3}\includegraphics[width=4.cm,clip]{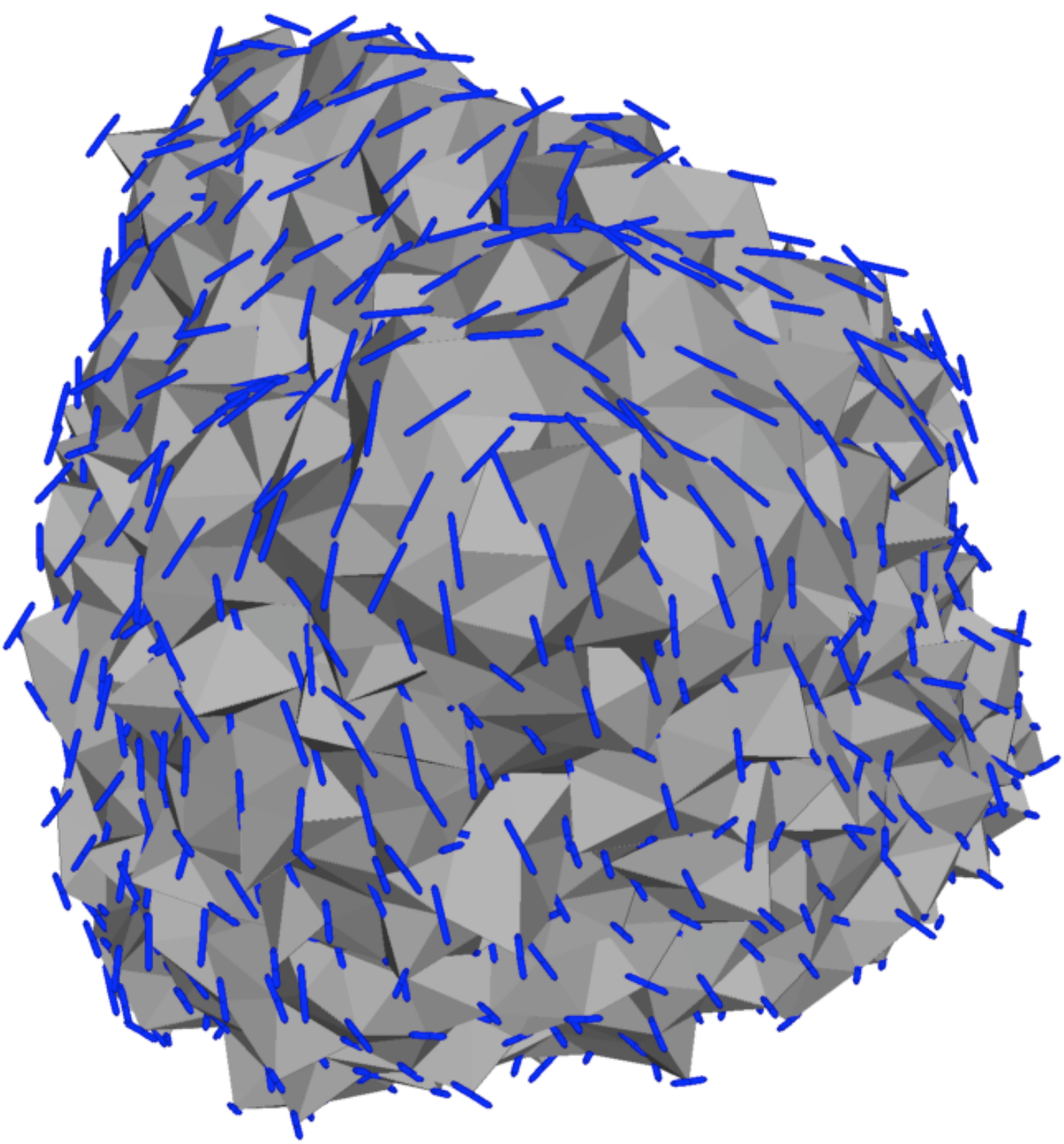}{{\large{\textbf{ (c)}}}}} 
\caption{ (Color online) Equilibrium configuration of a nematic embedded vesicle with $\kappa=0$, $c^0_{\parallel}=0$, $\epsilon_{LL}=3.0$, $\kappa_{\perp}=0 $ and
(a) $\kappa_{\parallel}=\,0$,  mere presence of an nematic field in the ordered phase cuts off the entropy dominated branched polymer phase seen otherwise (b) $\kappa_{\parallel}=20$ and (c) a corner with defect of index $+\frac{1}{2}$ is shown for $\kappa_{\parallel}=0$. All data are for a triangulated surface with 1202 vertices.}
\label{fig:all0}
\end{figure*}
\subsection{Membranes with in-plane nematic order}
We will consider the case of a randomly triangulated surface with an in-plane nematic field. 
These systems, in the continuum limit,  are described by a free energy functional which contains, in addition to the basic Helfrich curvature elastic part Eq.\eqref{helfrichham},  terms describing nematic-nematic interactions Eq.\eqref{Eq:nem-nem} and the coupling of the 
nematic field to the membrane curvature Eq.\eqref{Eq:nem-cur}.  For the discretized nematic-nematic interactions we have employed 
the Lebwohl-Lasher\cite{lebwohl_1972, fabbri_1986} model, described in Eq.\eqref{Eq:lebwohl}, which corresponds the one constant approximation of Frank's free energy given in Eq.({\eqref{Eq:nem-nem}).
 The total  discretized free energy functional then takes the form 
\begin{eqnarray}
{\cal H} & = & \frac{\kappa}{2}\sum_{v=1}^{N}{M(v)}^2  A(v) \nonumber \\
         &-& \frac{\epsilon_{LL}}{2} \sum_{v} \sum_{v^{'} \in \{v\}}  \left\{ \frac{3}{2} \cos^2(\phi(v,v^{'}))-\frac{1}{2} \right\} \nonumber \\
             &+&\frac{\kappa_{\parallel}}{2} \sum_{v=1}^{N}\left[H_{\hat{n}(v),\parallel}-c^{\parallel}_0\right]^2 A(v) \nonumber \\
             &+& \frac{\kappa_{\perp}}{2} \sum_{v=1}^{N}\left[H_{\hat{n}(v),\perp}-c^{\perp}_0\right]^2 A(v),  \label{eq:model} 
\end{eqnarray}  
where, $H_{\hat{n}(v),\parallel}= {n_1(v)}^2 c_{1}(v)+{n_2(v)}^2 c_{2}(v) $ and $H_{\hat{n}(v),\perp}={n_1(v)}^2 c_{2}(v)+{n_2(v)}^2 c_{1}(v)$ are the directional curvatures at a vertex $v$, see Eq.\eqref{eq:sectcurvature}. $M(v)=[c_1(v)+c_2(v)]/2$ is the corresponding mean curvature. Note that this free energy is expressed in the local  Darboux frame of reference, described in Sec. ~\ref{exp_obs}.  
$n_{1}(v)$ and $n_{2}(v)$ are the components of the nematic director  in this local frame, and $c_{1}(v)$ and $c_{2}(v)$ are its principal  curvatures. $A(v)$ is the area of the polygonal surface defined by its nearest neighbors.

We will, in what follows, demonstrate the use of the algorithm by studying the  effect of in-plane orientational ordering on membrane conformations. 
 A detailed quantitative analysis and phase diagram of the vesicles shapes and in-plane ordering that can result from Eq.(\ref{eq:model}) will be published elsewhere.
\subsubsection{Membrane stiffness originating from the nematic field}
 First we consider the case with  $\kappa=0$, $\kappa_{\parallel}\neq 0$ and $c_{\parallel}=0$. We choose  $\kappa_{\perp}=0$ so that  the nematic field does not directly influence the bending modulus perpendicular to it. Such a situation may arise in the case of long thread like inclusions.   $\epsilon_{LL}=3$ is chosen to favor in-plane nematic order.
  
  Characteristic equilibrium configurations corresponding to 
$\kappa_{\parallel}=0$  and $20$ are shown in  Fig.~\ref{fig:all0}. For $\kappa_{\parallel}=0$ the common shapes are deformed tetrahedrons
 with four well-defined corner points. The in-plane orientational field displays perfect nematic ordering except at the corner points where
a disclination with index $1/2$ is located.  A snapshot of one of these disclinations is shown in Fig.~\ref{fig:all0},c. Since these are the only disclinations, the total index is $2$, 
in accordance with Poincare's index theorem. The surface appear crinkled with local scale roughness. 
For $\kappa_{\parallel}=20$ the vesicle shape  becomes elongated, with the long axis following the orientation of the nematic field, with
sharp ends. The two $1/2$ defects  are now joined to form a defect of index $1$, and is located at the two ends.

Membrane without stiffness and nematic degrees of freedom has branched polymer configurations. While our simulations show that,  for the same system size,   such a phase is absent in membranes with in-plane order. It thus follows that in-plane ordering induces configurational stiffness of vesicles.  Signature of this stiffness can also be seen in the  distribution of eigenvalues of the gyration tensor for two different values of $\epsilon_{LL}$.  As can be seen in Fig.~\ref{fig:eigendist}, the distribution  of higher eigenvalues are  narrower for higher  $\epsilon_{LL}$, indicating stiffening.
\begin{figure}[!h]
\centering
\includegraphics[width=7.75 cm,clip]{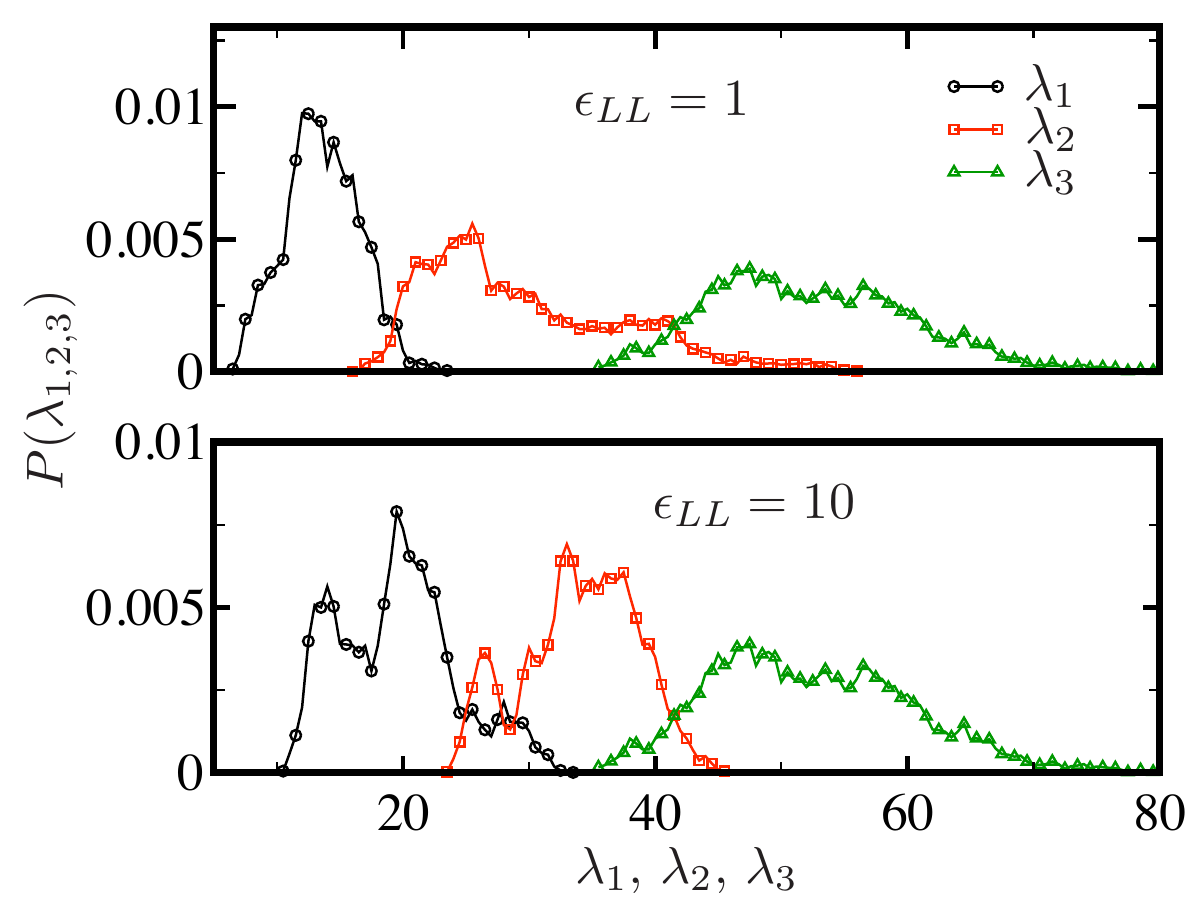}
\caption{(Color online) Distribution of the eigenvalues($\lambda_i$) of the gyration tensor, such that  $\lambda_1 < \lambda_2 < \lambda_3$,  for  $\kappa = \kappa_{\parallel}=\kappa_{\perp}=0$,  (a) $\epsilon_{LL}=1 $ and (b) $\epsilon_{LL}=10$.}
\label{fig:eigendist}
\end{figure}
  We note that the  anisotropic elasticity of the membrane, arising  through this  nematic orientation,  is  similar to that suggested by Fo\v{s}nari\v{c} {\em et al.}~\cite{Miha_2005}.
\subsubsection{Positive spontaneous curvature}
Making $c_0^{\parallel} > 0$ imply that  the nematic field  favors a  specific value of positive  curvature  along  the direction of its axis.
In Fig.~\ref{fig:pos-scur} is shown representative equilibrium configurations for $\epsilon_{LL}=3$, $\kappa_{\perp}=0 $, $\kappa_{\parallel}=20$, $c^0_{\parallel}=0.5$ and $\kappa=2.5({\rm a}),\ \ 10 ({\rm b})$. For $\kappa=2.5$ the vesicle shape transforms to branched structure with long irregular tubes of varying radius. The nematic field now spirals around the tubes.
The  angle made by the nematic field with the azimuthal direction  increases with  decrease in local tube radius. The caps of the
tubular structures are quipped with  disclination  pairs of index $1/2$, while  two disclinations with index $-1/2$ are situated in the branchpoints. The tubes themselves tend to spiral over longer distances, as can be seen from Fig.~\ref{fig:pos-scur},a.  This spiraling can  both be right and left handed, indicating no chiral preference.  For $\kappa= 10$ this picture persists, except the nematic ordering match up with the azimuthal direction of the tubes, no chiral ordering of the tubes are observed and the tube radius match with the that set by $c^0_{\parallel}$.    
\begin{figure}[]
\centering
{\bf
\subfloat{\label{possp-f1}\includegraphics[width=6.cm,clip]{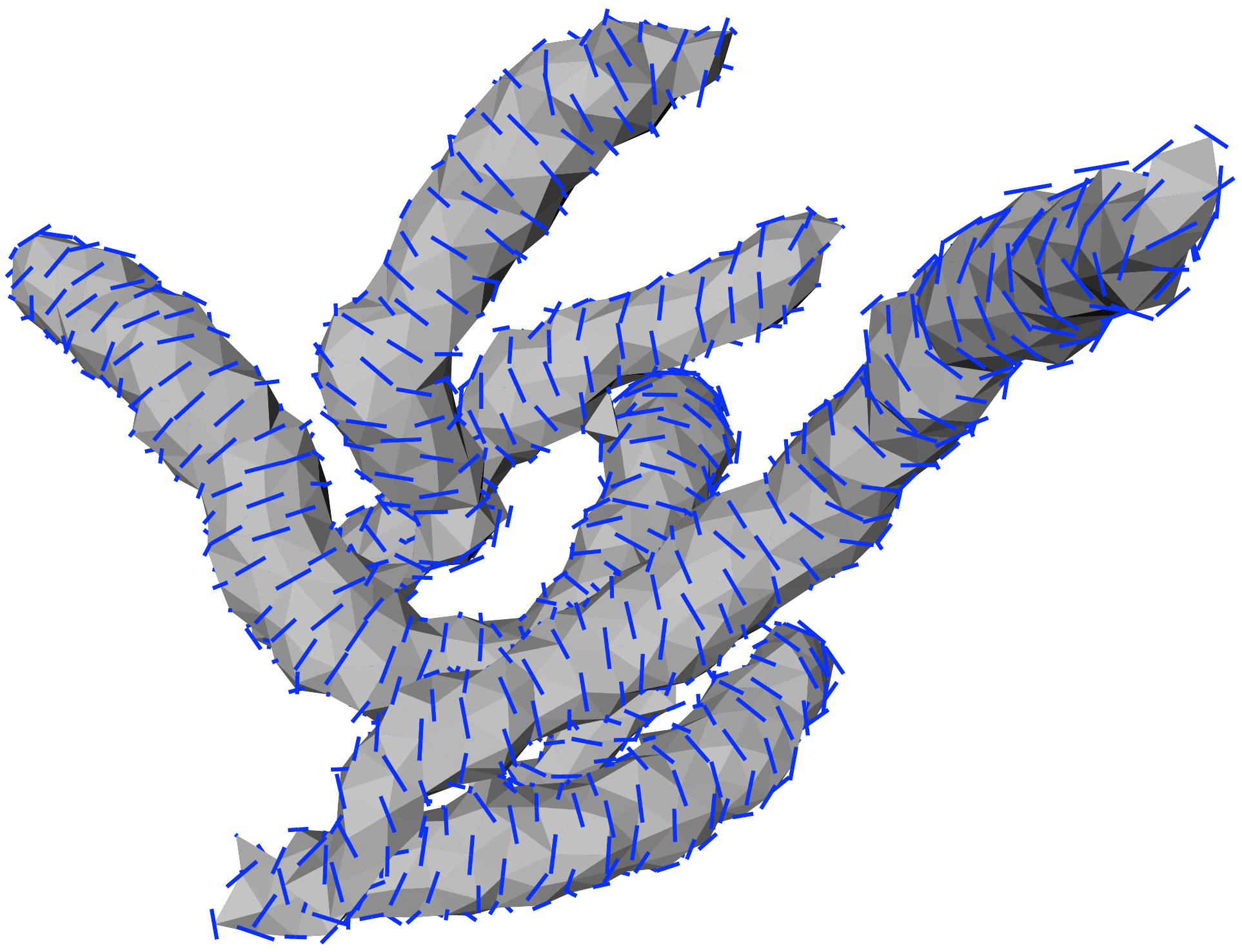}{{\large{\textbf{ (a)}}}}}\\  
\subfloat{\label{possp-f2}\includegraphics[width=6.cm,clip]{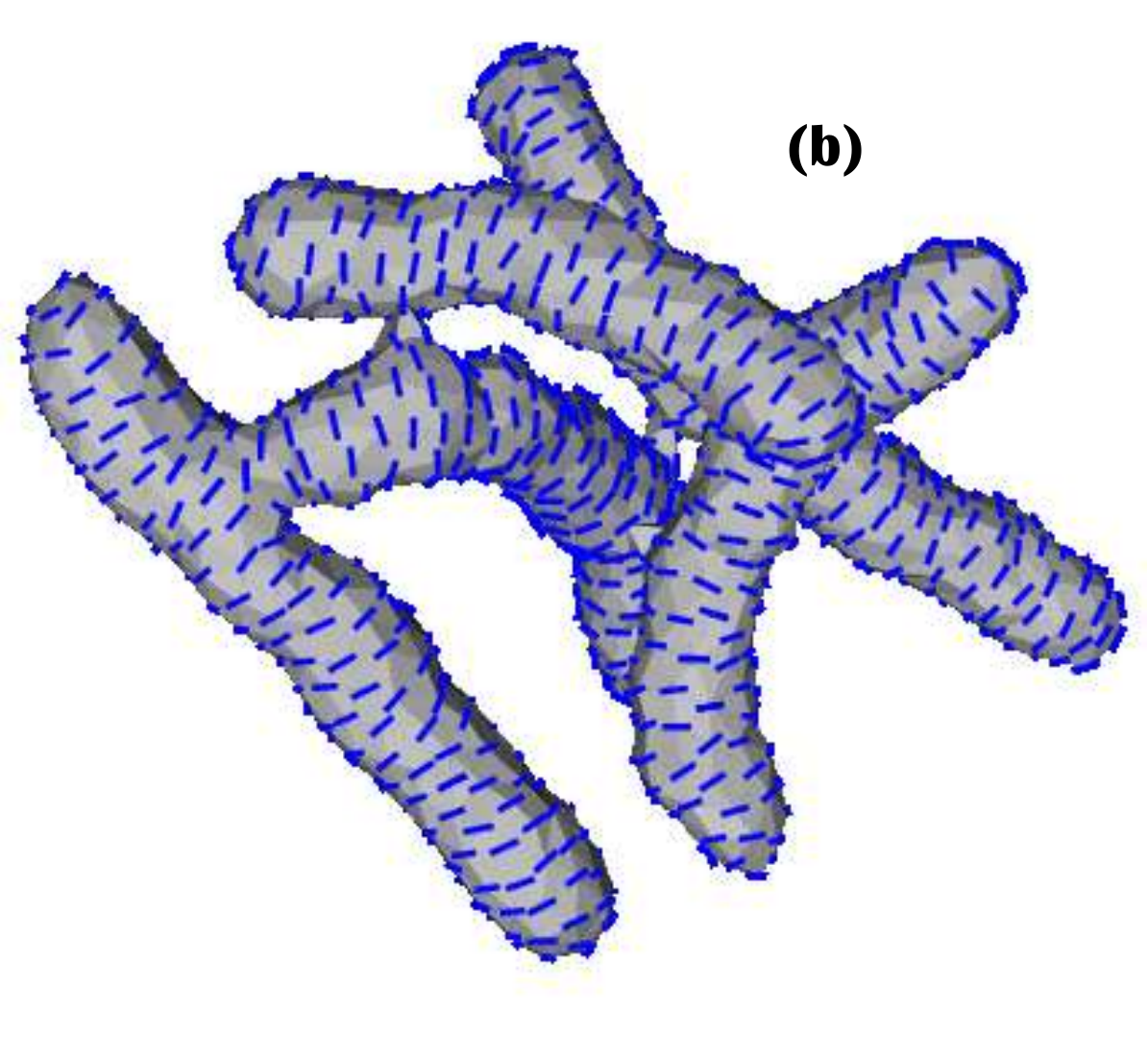}{{\large{\textbf{ (b)}}}}}}
\caption{ (Color online) Configurations of membranes, with $\kappa_{\parallel}=20$, $c^0_{\parallel}=0.5$,$\epsilon_{LL}=3$, $\kappa_{\perp}=0 $ 
for (a) $\kappa=2.5$ and (b)$\kappa=10$}
\label{fig:pos-scur}
\end{figure}
\subsubsection{ Negative spontaneous curvature}
Negative spontaneous curvature,  $c_0^{\parallel} <  0$,  implies that the nematic will now prefer to orient along directions where the membrane curvature is negative ( curved into the vesicle).
In Fig.~\ref{fig:neg-scur} is shown examples of equilibrium configurations for $\kappa =0$ and $\kappa=10$, where 
 $\epsilon_{LL}=3.0$, $\kappa_{\parallel}=30$, $\kappa_{\perp}=0 $, $c^0_{\parallel}=-0.5$ and $\kappa_{\perp}=0 $.   Inward tubulation results in stiffening of the outer boundary of the vesicle as shown in Fig.~\ref{fig:neg-scur}. In contrary to the tubulation seen in the case of $C_0^{\parallel}>0$, self avoidance condition of the membrane now prevents complete tube formation.
 Similar to the  $c_0^{\parallel} > 0$ case, increasing $\kappa$ increases the thickness of the tubes. The nematic ordering is along the
 azimuthal direction for $\kappa=10$, while spiraling is observable for $\kappa=0$. On the outer surface, defects of  index $-\frac{1}{2}$ are clearly observed.
\begin{figure*}
\centering
\subfloat{ {{\large{\textbf{ (a)}}}}\label{negsp-f1}\includegraphics[width=4.1cm,clip]{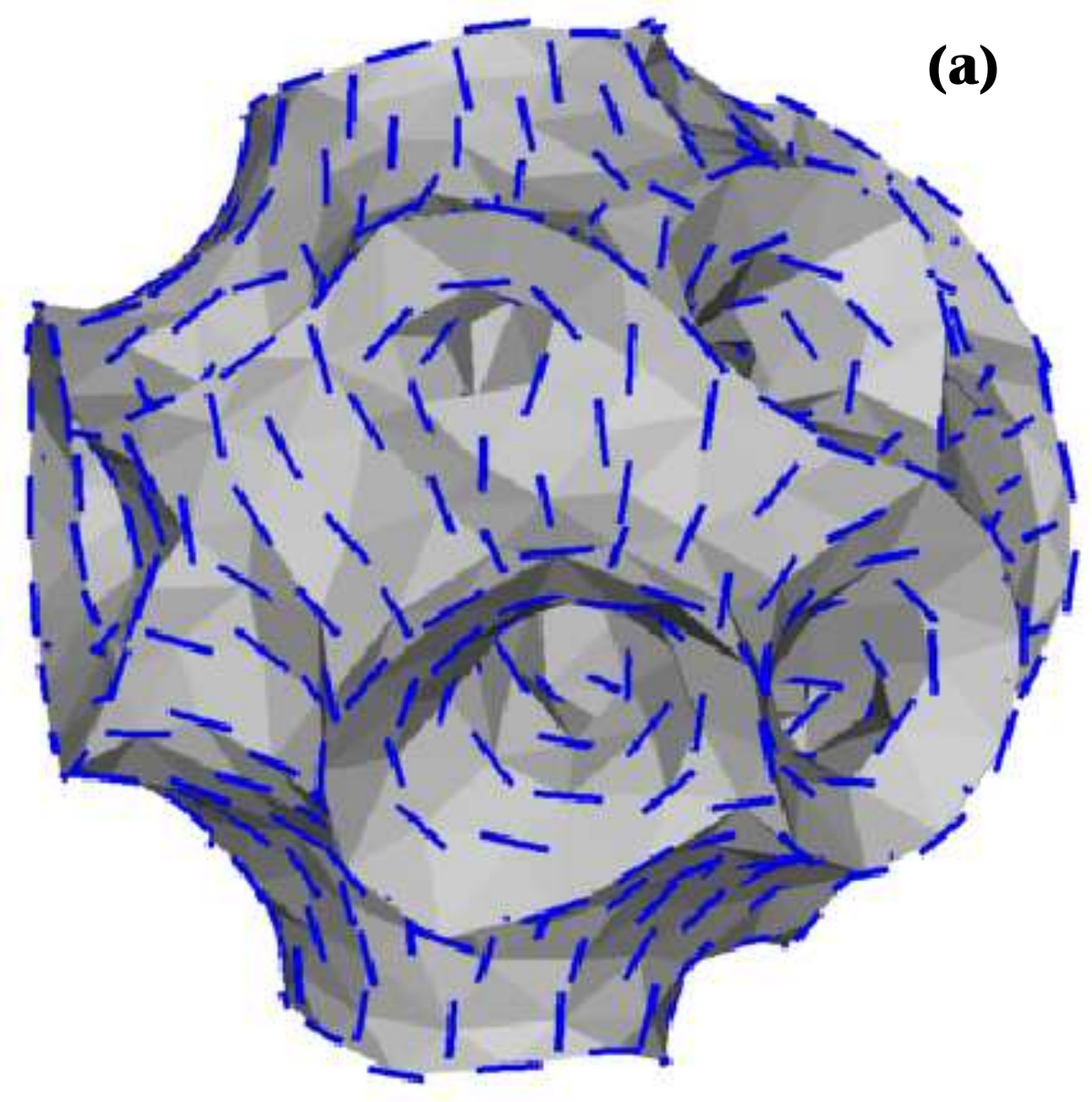}}   
\subfloat{{{\large{\textbf{ (b)}}}}\label{negsp-f2}\includegraphics[width=4.1cm,clip]{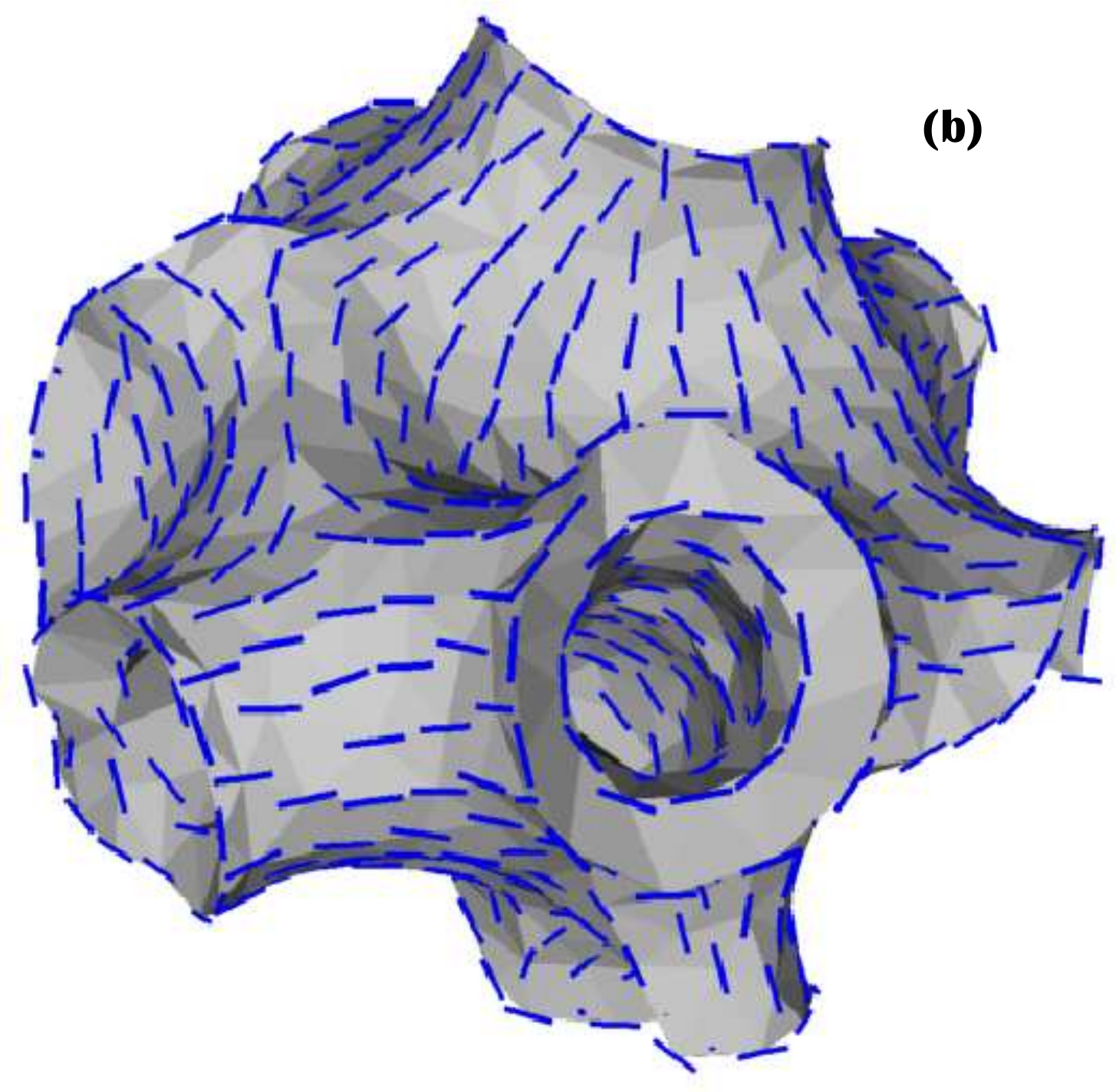}}
\subfloat{{{\large{\textbf{ (c)}}}}\label{negsp-f3}\includegraphics[width=4.1cm,clip]{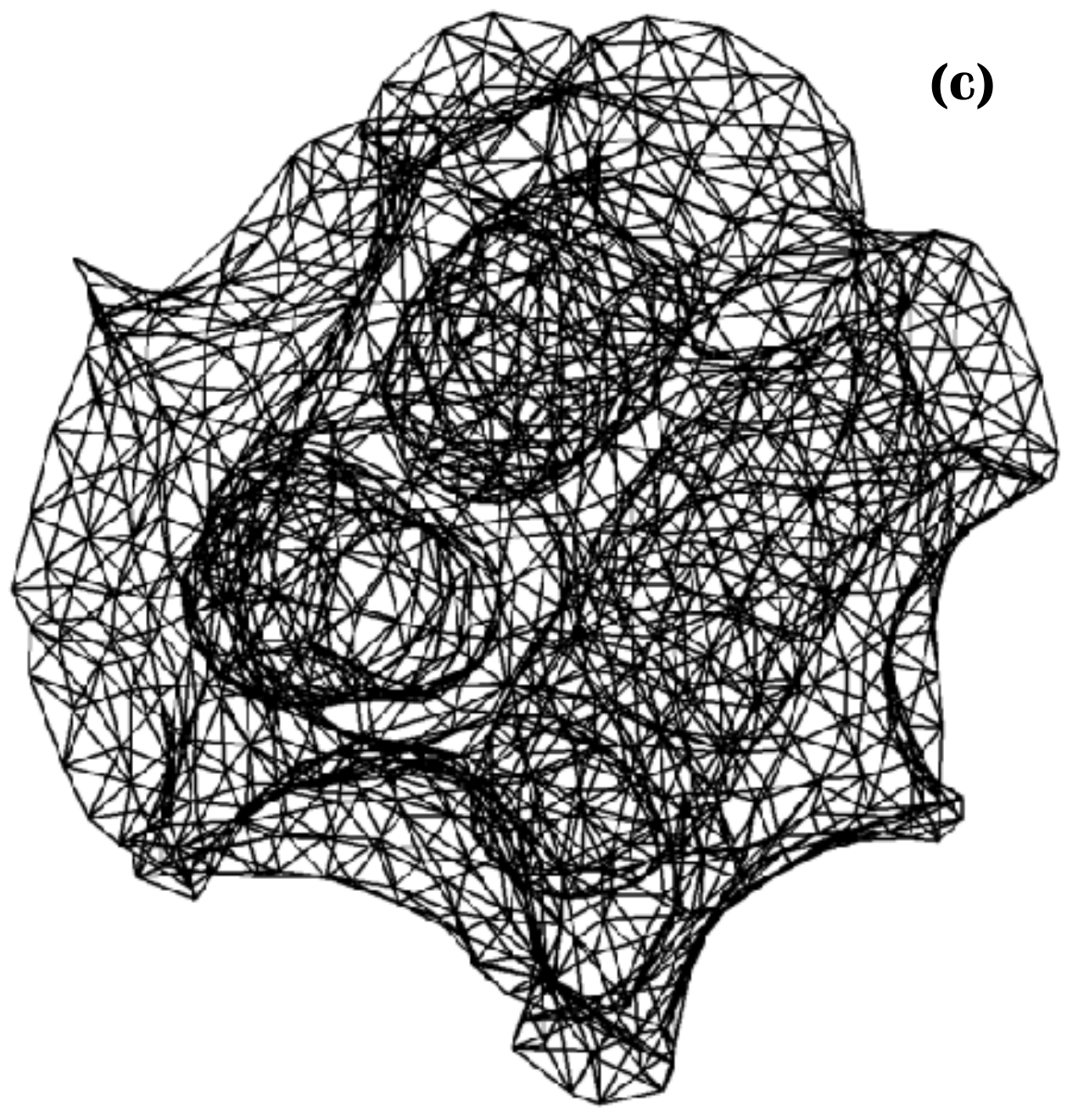}}   
\caption{ (Color online) Configurations of membranes, with $\kappa_{\parallel}=30$, $\epsilon_{LL}=3.0$, $c^0_{\parallel}=-0.5$, $\kappa_{\perp}=0 $ for (a)\,$\kappa=0$ and (b)\,$\kappa=10$. (c) is the mesh representation of the surface in  (b) which clearly shows tubes grown inwards.} 
\label{fig:neg-scur}
\end{figure*}


\section{Conclusion} \label{conclusion}
We have presented  a methodology for calculating surface quantifiers on a self-avoiding triangulated random surface models of fluid membranes.
The method involves calculations of the local geometrical properties at the vertex positions of the surface, 
e.g., calculation of the local Darboux frame and the principal curvature radii of the surface.  
We have described  a procedure for parallel transport of in-plane vectors between vertex points.  
 We have implemented the numerical model and performed Monte Carlo simulations of the equilibrium properties of the surface. 
The simulations of the discretized form  for the  Helfrich's free energy  are in   good qualitative agreement with the results from 
previous numerical simulations. 
In the flexible limit of low bending rigidity the membrane scales as a branched polymer and a scaling relation involving
volume, system size and persistence length holds.  For small values of $\kappa$,  calculations using the new discrete Hamiltonian shows  a faster increase, as a function of $\kappa$, in the persistence length compared to the previous model.

The model has been extended to include  an in-plane  nematic field and   equilibrium shapes have been obtained for some  simple examples. We show that the presence of a nematic ordering leads to suppression of the branched polymer phase even when the bare bending rigidity is zero. 
 The conformational changes in a fluid membrane brought about by the anisotropy in the bending rigidity½ and the spontaneous 
curvature induced by the nematic  field  are demonstrated.   We have demonstrated that the presence of the in-plane nematic field leads
to coupling between  geometry and nematic defect structures of the membrane. It is shown that this coupling can lead to chiral structures   in membrane even in the absence of explicit chiral terms in the Hamiltonian.

\section*{Acknowledgements}
The MEMPHYS-Center for Biomembrane
Physics is supported by the Danish National Research Foundation. Computations were carried out at the HPC facility at IIT Madras and    Danish Center of Scientific Computing  at SDU. 


\appendix
\section{Householder Transformation}
Consider two orthonormal frames of reference given by the coordinates $(\hat{x},\hat{y},\hat{z})$ and $(\hat{a},\hat{b},\hat{c})$.
The Householder matrix $H$, can be used to rotate $\hat{z}$ in frame 1 to $\hat{c}$ in frame 2, such that $(\hat{x},\hat{y})$ now 
are some  arbitrary vectors in the plane formed by $(\hat{a},\hat{b})$. Define a vector,
\begin{equation}
W=\frac{\hat{x} \pm \hat{c}}{|\hat{x} \pm \hat{c}|}
\end{equation}
with a minus sign if $||\hat{x}-\hat{c}||\,>\,||\hat{x}+\hat{c}||$ and a plus if otherwise.The Householder matrix is then defined as,

\begin{equation}
H=\mathbbm{1}-2WW^\dagger
\end{equation}

\end{document}